\documentclass[noccpublish,nocclayout,nocctextarea,noccsectioning,letterpaper]{cc}
\usepackage[scale={.753, .8107},marginratio={1:1, 3:4}]{geometry}
\usepackage[font=small,margin=12pt,skip=9.7pt]{caption}

\title{Lipschitz Continuous Ordinary Differential Equations are 
       Polynomial-Space Complete}

\author{Akitoshi Kawamura}

\date{April 2010}

\usepackage{graphicx}

\newcommand{\kome}{\textup{($*$)}}
\newcommand{\classP}{\mathbf P}
\newcommand{\classPSPACE}{\mathbf{PSPACE}}

\newcommand{\classNumberP}{\mathbf{\#P}}
\newcommand{\classNP}{\mathbf{NP}}
\newcommand{\classEXP}{\mathbf{EXPTIME}}
\newcommand{\classEXPSPACE}{\mathbf{EXPSPACE}}
\newcommand{\Nset}{\mathbf N}
\newcommand{\Zset}{\mathbf Z}
\newcommand{\Qset}{\mathbf Q}
\newcommand{\Rset}{\mathbf R}
\renewcommand{\d}{\mathrm d}
\newcommand{\tcolon}{\colon}
\newcommand{\tto}{\mathbin\to}

\newcommand{\probQBF}{\textsc{qbf}}

\begin{document}

\maketitle

\begin{abstract}
In answer to Ko's question raised in 1983, we show that an initial value problem given by a polynomial-time computable, Lipschitz continuous function can have a polynomial-space complete solution.  The key insight is simple: the Lipschitz condition means that the feedback in the differential equation is weak.  We define a class of polynomial-space computation tableaux with equally weak feedback, and show that they are still polynomial-space complete.  The same technique also settles Ko's two later questions on Volterra integral equations. 

\medskip

\noindent\textbf{Keywords:} 
computable analysis; computational complexity; initial value problem; Lipschitz condition; ordinary differential equations; Picard--Lindel\"of Theorem; polynomial space.
% \begin{subject}
%   03F60, % Constructive and recursive analysis
%   68Q17, % Computational difficulty of problems (lower bounds, 
%          % completeness, difficulty of approximation, etc.)
%   65Y20, % Complexity and performance of numerical algorithms
%   65L05, % Initial value problems
%   03D15 % Complexity of computation
% \end{subject}
\end{abstract}

\section{Introduction}

  Let $g \tcolon [0, 1] \times \Rset \tto \Rset$ be a continuous function and 
  consider the initial value problem
\begin{align}
 \label{equation: problem}
  h (0) 
&
 =
  0, 
&
  h' (t)
&
 =
  g \bigl( t, h (t) \bigr), 
\quad
  t \in [0, 1]. 
\end{align}
  A well-known sufficient condition 
  (see the beginning of \ref{section: question} for a proof sketch) for 
  this equation to have a unique solution $h \tcolon [0, 1] \tto \Rset$
  is that $g$ be
  \emph{Lipschitz continuous} (in its second argument), 
  i.e., 
\begin{equation}
 \label{equation: Lipschitz}
   \lvert g (t, y _0) - g (t, y _1) \rvert 
  \leq 
   Z \cdot \lvert y _0 - y _1 \rvert, 
  \qquad
   t \in [0, 1], \ y _0, y _1 \in \Rset
\end{equation}
  for some constant $Z$ 
  independent of $y _0$, $y _1$ and $t$. 
  We are interested in the computational complexity 
  of the solution~$h$ under this condition. 

  Our model of computation of real functions, 
  which will be reviewed in \ref{section: computable analysis}, 
  is adopted from computable analysis
  and is thus consistent with the conventional notion of computability. 
  We formulate our main result in 
  \ref{section: question}: 
  the solution~$h$ of the above equation can be polynomial-space complete, 
  even if $g$ is polynomial-time computable. 
  This was open since 
\cite{ko83:_comput_compl_of_ordin_differ_equat}. 
  The essential part of the proof is given in \ref{section: main proof}, 
  where we construct a certain family of real functions
  that can be used as building blocks for the desired $g$ and $h$. 
  The main idea is to regard the differential equation with the Lipschitz condition
  as a polynomial-space computation tableau with some restrictions. 
  In \ref{section: variants}, 
  we state a few variants of the main theorem, 
  two of which solve the problems about Volterra integral equations 
  posed by 
\cite{ko92:_comput_compl_of_integ_equat}. 
  These variants are also proved using the same building blocks, 
  as shown in \ref{section: proofs}. 
  \ref{section: related} discusses related results and open problems. 

\section{Computational complexity of real functions}
 \label{section: computable analysis}

  The study of mathematical analysis from the viewpoint of computability 
  is called \emph{computable analysis}; 
\cite{brattka08:_tutor_comput_analy}
  and
\cite{weihrauch00:_comput_analy} 
  are good introductions to the field. 
  We review the basic definitions 
  brief\textcompwordmark ly here, 
  refining them for our complexity consideration where necessary. 

  The computability notion for real functions 
  equivalent to ours dates back at least to 
\cite{grzegorczyk55:_comput_funct}. 
  Polynomial-time computability of real functions
  was introduced by 
\cite{ko82:_comput_compl_of_real_funct} 
  using oracle machines, 
  and is equivalent, at least in our context, to the one defined 
  by the type-two machine and the signed digit representation 
  (Chapter~7 of 
\cite{weihrauch00:_comput_analy}). 

\subsection{Computing real functions}
 \label{subsection: computability}

  Since real numbers cannot be encoded into strings, 
  we encode them into functions from strings to strings. 
  We say that a real number~$t$ is 
\emph{represented}
  by a string function~$A$ if 
  for any $m \in \Nset$, 
  the string $A (0 ^m)$ 
  is the binary notation (with a sign bit at the beginning) of either 
  $\lfloor 2 ^m t \rfloor$ or 
  $\lceil 2 ^m t \rceil$, 
  where $\lfloor \mathord\cdot \rfloor$ and $\lceil \mathord\cdot \rceil$ mean
  rounding down and up to the nearest integer, respectively. 
  In effect, $A (0 ^m)$ gives an 
  approximation of $t$ with precision~$2 ^{-m}$
  by a multiple of $2 ^{-m}$. 
  We also say that $A$ is a 
\emph{name}
  of $t$. 

  Computation of real functions is 
  realized by \emph{oracle Turing machines}
  (henceforth just \emph{machines}) working on 
  such names~$A$. 
  In addition to the input, output and work tapes, 
  the machine has a query tape 
  and can consult an external oracle~$A$ by 
  entering a distinguished state; 
  the string~$v$ which is on the query tape at this moment
  is then replaced by $A (v)$ in one step. 
  We write $M ^A$ for 
  the string-to-string function computed by machine~$M$ with oracle~$A$. 

\begin{definition}
 \label{definition: polynomial-time computable}
  A machine~$M$ 
\emph{computes}
  a function $
f \tcolon [0, 1] \tto \Rset
  $ if
  for any $t \in [0, 1]$ and 
  any name $A$ of it, 
  $M ^A$ is a name of $f (t)$. 
\end{definition}

\begin{figure}
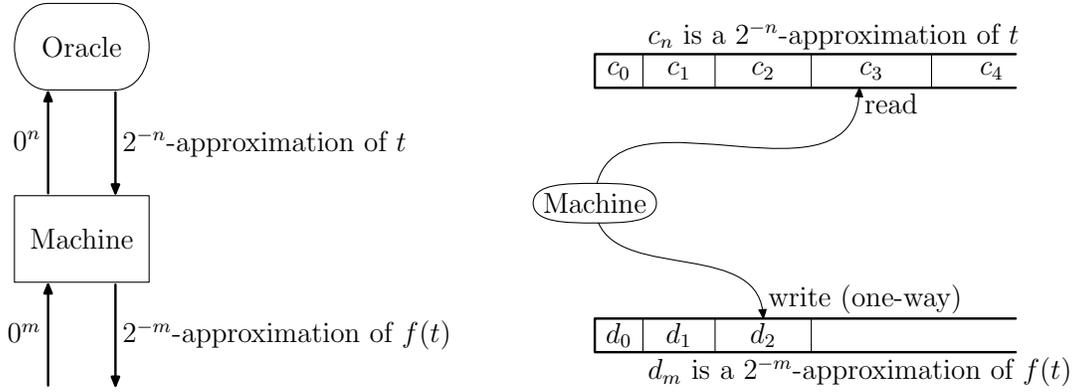

\begin{center}
\hfill
 \includegraphics[clip,scale=.91]{machine_2_cm.eps}%
\hfill
 \includegraphics[clip,scale=.91]{machine_1_cm.eps}%
\hfill\mbox{}
 \caption{%
  To compute a real function $f$, 
  the machine 
  should output an approximation of $f (t)$ with given precision~$2 ^{-m}$
  by consulting the oracle for 
  approximations of $t$ with any precision~$2 ^{-n}$ it desires (left). 
  An alternative picture (right) is that the machine converts
  any stream of improving approximations of $t$
  to a stream of improving approximations of $f (t)$. 
 }
 \label{figure: machine}
\end{center}
\end{figure}

  Thus, computation of a real function $f$ 
  is a 
  Turing reduction of 
  (a name of) $f (t)$
  to $t$ (\ref{figure: machine}, left). 
  A little thought shows that 
  it can equivalently be visualized as 
  a Turing machine that, 
  given on the input tape an infinite sequence of approximations of $t$, 
  writes approximations of $f (t)$
  endlessly on the one-way output tape~%
  (\ref{figure: machine}, right). 

  A machine runs in \emph{polynomial time} if 
  there is a polynomial $p \tcolon \Nset \tto \Nset$ such that, 
  for any input string~$u$, 
  it halts within $p (\lvert u \rvert)$ steps
  regardless of the oracle. 
  A real function is (\emph{polynomial-time}) \emph{computable} if 
  some machine (that runs in polynomial time) computes it. 

  When writing an approximation of $f (t)$ with precision $2 ^{-m}$, 
  the machine knows $t$ only to some finite precision $2 ^{-n}$. 
  Hence, all computable functions are continuous. 
  If the machine runs in polynomial time, 
  then this $n$ is bounded polynomially in $m$. 
  Hence, all polynomial-time computable functions~$f$ have 
  a polynomial \emph{modulus of continuity} (\ref{figure: modulus}): 
  there is a polynomial $p$ such that $
\lvert f (t _0) - f (t _1) \rvert < 2 ^{-m}
  $ for all $t _0$, $t _1 \in [0, 1]$ and $m \in \Nset$ 
  with $\lvert t _0 - t _1 \rvert < 2 ^{-p (m)}$ 
  (note that in our setting it makes sense to put $p$ in the exponent, 
  deviating from some authors' terminology 
  where a modulus of continuity means 
  a function that takes an upper bound on $\lvert t _0 - t _1 \rvert$ to 
  that on $\lvert f (t _0) - f (t _1) \rvert$). 
  In fact, it is not hard to see that 
  polynomial-time computability can be characterized by 
  this plus the assertion that $f$ can be approximated at rationals: 
\begin{figure}
\begin{center}
 \includegraphics[clip,scale=.91]{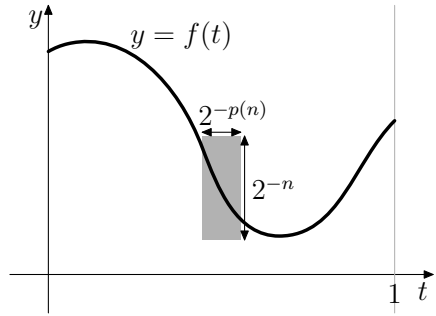}%
 \caption{Modulus of continuity~$p$.}
 \label{figure: modulus}
\end{center}
\end{figure}

\begin{lemma}
 \label{lemma: polynomial modulus}
  A function~$f \tcolon [0, 1] \tto \Rset$ is polynomial-time computable 
  if and only if 
  it has a polynomial modulus of continuity and 
  there is a polynomial-time computable function $
g \tcolon ([0, 1] \cap \Qset) \times \{0\} ^* \tto \Qset
  $ such that 
\begin{equation}
 \lvert g (d, 0 ^n) - f (d) \rvert < 2 ^{-n}, 
\qquad
 d \in [0, 1] \cap \Qset, \ n \in \Nset, 
\end{equation}
  where rational numbers are 
  encoded in a reasonable way (e.g., using fractions 
  whose numerator and denominator are integers written in binary). 
\end{lemma}

  Many familiar continuous functions are computable. 
  For example, 
  it is easy to see that 
  the sine function restricted to $[0, 1]$ 
  is polynomial-time computable, 
  because an approximation of
\begin{equation}
  \sin t = t - \frac{t ^3}{3!} + \frac{t ^5}{5!} - \frac{t ^7}{7!} + \cdots 
\end{equation}
  to precision $2 ^{-m}$ can be found by 
  approximating the sum of polynomially many (in $m$) initial terms, 
  since this series converges fast enough on $[0, 1]$. 

  The above definition can be straightforwardly extended 
  to functions on compact intervals other than $[0, 1]$
  and on $d$-dimensional rectangles (by considering machines taking $d$ oracles). 
  Also, polynomial-space, exponential-time and exponential-space computability 
  is defined analogously to polynomial-time computability, 
  where by ``exponential'' we mean $2 ^{n ^{\mathrm O (1)}}$ 
  (and not $2 ^{\mathrm O (n)}$). 
  Here, 
  we count the query tape in 
  when defining space complexity. 
  The definition in Section~7.2.1 of 
\cite{ko91:_comput_compl_of_real_funct}
  states to the contrary, 
  but his subsequent theorems 
  build on the definition that does charge the query tape
  (on the other hand, 
  his argument in Chapter~4 that 
  the query tape should not be counted 
  in discussing logarithmic space is correct). 

\subsection{Completeness}

  We now introduce terminology to state our main results 
  which say that certain real functions are ``hard'' to compute. 
  We regard a language~$L$ as a set of strings 
  or as a $\{0, 1\}$-valued function interchangeably, 
  so that $L (u) = 1$ means $u \in L$. 
  We write $\classP$, $\classNP$, $\classPSPACE$, $\classEXP$, $\classEXPSPACE$ for 
  the standard classes of languages
  and $\classNumberP$ for the function class; see
\cite{papadimitriou94:_comput_compl}. 

\begin{figure}
\begin{center}
 \includegraphics[clip,scale=.91]{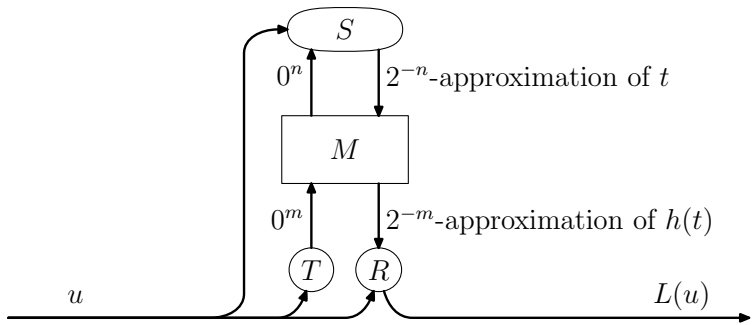}%
 \caption{$L$ reduces to $h$ via $R$, $S$ and $T$.
          This means that, 
          using a hypothetical machine~$M$ computing $h$ 
          (in the sense of \ref{definition: polynomial-time computable})
          as a black box, 
          we can compute $L$ in polynomial time
          in the way depicted above.}
 \label{figure: typetworeduction}
\end{center}
\end{figure}

\begin{definition}
 \label{definition: completeness}
  A function~$L$ (over strings) is said to 
\emph{reduce}
  to a real function $h \tcolon [0, 1] \tto \Rset$ if 
  the following holds for some 
  polynomial-time computable functions $R$, $S$, $T$: 
  Let $u$ be a string, and 
  suppose that the function 
  taking string~$v$ to $S (u, v)$ is a name of a real number $t \in [0, 1]$. 
  Then $L (u) = R (u, \psi (T (u)))$
  for any name~$\psi$ of $f (t)$
  (\ref{figure: typetworeduction}). 
\end{definition}

  For a complexity class~$\mathcal C$, 
  we say that a real function is 
\emph{$\mathcal C$-hard}
  if all problems in $\mathcal C$ reduce to it. 
  A real function is 
\emph{polynomial-space} 
(resp.\ \emph{exponential-space}) \emph{complete} 
  if it is 
  polynomial-space (resp.\ exponential-space) computable 
  and $\classPSPACE$-hard (resp.\ $\classEXPSPACE$-hard). 

The above definitions of reduction and completeness can be viewed as 
a special case of those by \cite{beame98:_relat_compl_of_np_searc_probl}, 
and are also consistent with 
Definition~2.2 of \cite{ko92:_comput_compl_of_integ_equat}. 

\section{Ko's question and our main result}
 \label{section: question}

  Now we return to the differential equation~\ref{equation: problem}. 
  The fact that \ref{equation: Lipschitz} guarantees 
  a unique solution 
  is known as (a variant of) the Picard--Lindel\"of (or Cauchy--Lipschitz) Theorem, 
  and can be proved as follows. 
  Let $C$ be the set of all continuous real functions on $[0, 1]$. 
  A solution of \ref{equation: problem} is 
  a fixed point of the operator $T \tcolon C \tto C$ defined by 
\begin{align}
 T (h) (t) & = \int _0 ^t g \bigl( \tau, h (\tau) \bigr) \, \d \tau, 
& 
 t \in [0, 1].
\end{align}
  The existence and uniqueness of this fixed point
  follow from the contraction principle (Banach fixed point theorem), 
  because a simple calculation shows that, 
  for the metric~$d$ on $C$ given by $
  d (h _0, h _1) 
 = 
  \max _{t \in [0, 1]} \exp (-2 Z t) \lvert h _0 (t) - h _1 (t) \rvert
  $, we have
  $d (T (h _0), T (h _1)) \leq d (h _0, h _1) / 2$. 

  We assume the following, 
  and ask how complex $h$ can be: 
\begin{itemize}
 \item[\kome]
  $g \tcolon [0, 1] \times \Rset \tto \Rset$ and $h \tcolon [0, 1] \tto \Rset$
  satisfy \ref{equation: problem}, 
  $g$ satisfies \ref{equation: Lipschitz}, 
  and $g$ is polynomial-time computable.
\end{itemize}
  Strictly speaking, 
  we have defined polynomial-time computability 
  only for functions on a compact rectangle. 
  What we mean here is 
  that the restriction of $g$ to $[0, 1] \times [\min h, \max h]$, say, is 
  polynomial-time computable. 
  Equivalently, 
  we could write $g \tcolon [0, 1] \times [-1, 1] \to \Rset$ and 
  add the clause ``the values of $h$ stays within $[-1, 1]$''
  to \kome{}; this does not essentially change our result, because 
  we can always make $h$ stay within $[-1, 1]$ 
  by scaling $g$ and $h$ down by a constant factor, 
  which does not affect polynomial-time computability. 
  There is a way to extend
  \ref{definition: polynomial-time computable} 
  to functions with unbounded domain, 
  as in \cite{hoover90:_feasib_real_funct_and_arith_circuit}
  or pp.\thinspace57--58 of \cite{ko91:_comput_compl_of_real_funct}, 
  but we choose our simpler definition. 

  As 
\cite{ko83:_comput_compl_of_ordin_differ_equat} 
  points out
  by analyzing the Euler method, 
  \kome{} implies that $h$ is polynomial-space computable. 
  From this it follows (Lemma~2.2 of 
\cite{ko83:_comput_compl_of_ordin_differ_equat})
  that 
  if $\classP = \classPSPACE$, 
  then \kome{} implies that $h$ is polynomial-time computable. 
  We will prove a lower bound that matches this upper bound: 

\begin{theorem}
 \label{theorem: main}
  There are functions $g$ and $h$ satisfying~\kome{} 
  such that $h$ is $\classPSPACE$-hard (and thus polynomial-space complete). 
\end{theorem}

\begin{corollary}
 \label{corollary: answer to Ko}
  $\classP = \classPSPACE$ if and only if 
  \kome{} always implies that $h$ is polynomial-time computable. 
\end{corollary}

  This solves the main problem left open in 
\cite{ko83:_comput_compl_of_ordin_differ_equat}. 
  He had proved there a partial result 
  essentially stating that \ref{theorem: main} holds true 
  if the Lipschitz condition~\ref{equation: Lipschitz} in 
  the assumption~\kome{} is 
  replaced by a weaker condition. 

  We remark that the special case of 
  the equation~\ref{equation: problem} where $g$ ignores its second argument 
  reduces to integration, 
  whose complexity is summarized as follows
  in the style similar to 
  \ref{theorem: main} and 
  \ref{corollary: answer to Ko}. 

\begin{theorem}[essentially by \cite{friedman84:_comput_compl_maxim_integ}]
  There are a polynomial-time computable function $g \tcolon [0, 1] \tto \Rset$ and 
  a $\classNumberP$-hard function $h \tcolon [0, 1] \tto \Rset$ such that
\begin{align}
 \label{equation: integration}
 h (t) & = \int _0 ^t g (\tau) \, \d \tau, &
 t \in [0, 1]. 
\end{align}
\end{theorem}

\begin{corollary}[\cite{friedman84:_comput_compl_maxim_integ}]
  $\classP = \classP ^{\classNumberP}$ if and only if 
  for all polynomial-time computable $g \tcolon [0, 1] \tto \Rset$, 
  the function $h$ 
  defined by \ref{equation: integration} 
  is polynomial-time computable. 
\end{corollary}

  The relation to the counting class is not surprising: 
  as is apparent from Friedman's proof, 
  the intuition behind this result is that 
  approximating the integral is 
  to \emph{count} the number of grid points below the graph of $g$. 

  This bound of $\classP ^{\classNumberP}$ has been the 
  best known lower bound also for our differential equation. 
  \ref{theorem: main} improves this to $\classPSPACE$. 

\section{Proof of the theorem}
 \label{section: main proof}

  We present the proof backwards, reducing \ref{theorem: main} to 
  \ref{lemma: main} and then reducing \ref{lemma: main} to \ref{lemma: discrete}. 
  In \ref{subsection: building blocks}, 
  we state \ref{lemma: main}
  asserting the existence of 
  a certain family of pairs of functions $(g _u) _u$ and $(h _u) _u$, 
  from which the functions $g$ and $h$ in \ref{theorem: main} can be constructed. 
  \ref{subsection: discrete ivp and the Lipschitz condition} 
  shows that \ref{lemma: main} follows from \ref{lemma: discrete}, 
  which asserts the $\classPSPACE$-completeness of
  a discrete version of the initial value problem. 
  This discrete problem is like a $\classPSPACE$ computation tableau, 
  but with a certain restriction similar to the Lipschitz condition. 
  \ref{subsection: discrete ivp is hard}
  then completes the proof by showing \ref{lemma: discrete}. 

\subsection{Building blocks}
 \label{subsection: building blocks}

  To state \ref{lemma: main}, 
  we need to extend the definition of computation 
  in \ref{subsection: computability}
  to \emph{families} of real functions indexed by strings~$u$. 
  This is done in the natural way, 
  by giving $u$ as another string input to the machine. 
  For example, a family~$(g _u) _u$ of functions $
g _u \tcolon [0, 1] \times [-1, 1] \tto \Rset
  $ is 
  computed by a machine~$M$ 
  if for any names $A$ and $B$ of $t \in [0, 1]$ and $y \in [-1, 1]$, 
  the function that takes string~$0 ^m$ to $M ^{A, B} (u, 0 ^m)$
  is a name of $g _u (t, y)$. 
  Note that in this case, 
  claiming that $M$ runs in polynomial time means that 
  it halts in time polynomial in $\lvert u \rvert + m$. 

\begin{lemma}
 \label{lemma: main}
  Let $L \in \classPSPACE$ and 
  let $\lambda \tcolon \Nset \tto \Nset$ be a polynomial. 
  Then there exist a polynomial $\rho \tcolon \Nset \tto \Nset$ 
  and families of functions $g _u \tcolon [0, 1] \times [-1, 1] \tto \Rset$ 
  and $h _u \tcolon [0, 1] \tto \Rset$ indexed by binary strings~$u$ 
  such that the family $(g _u) _u$ is polynomial-time computable 
  and for each $u$ we have
\begin{enumerate}
 \item \label{enumi: range}
  $h _u (t) \in [-1, 1]$ for all $t \in [0, 1]$; 
 \item \label{enumi: boundary}
  $g _u (0, y) = g _u (1, y) = 0$ for all $y \in [-1, 1]$; 
 \item \label{enumi: equation}
  $
  h _u (0) = 0
  $ and $
  h' _u (t)
 =
  g _u (t, h _u (t))
  $ for all $t \in [0, 1]$; 
 \item \label{enumi: Lipschitz}
  $
   \lvert g _u (t, y _0) - g _u (t, y _1) \rvert 
  \leq 
   2 ^{-\lambda (\lvert u \rvert)} \lvert y _0 - y _1 \rvert 
  $ for any $
t \in [0, 1]$ and $y _0, y _1 \in [-1, 1]
  $; 
 \item \label{enumi: output}
  $
 h _u (1) 
= 
 2 ^{-\rho (\lvert u \rvert)} L (u)
  $. 
\end{enumerate}
\end{lemma}

  We thus have a family of functions $g _u$ that each 
  give an initial value problem whose solution~$h _u$ 
  encodes $L (u)$ in its final value $h _u (1)$. 

\begin{figure}
\begin{center}
 \includegraphics[clip,scale=.91]{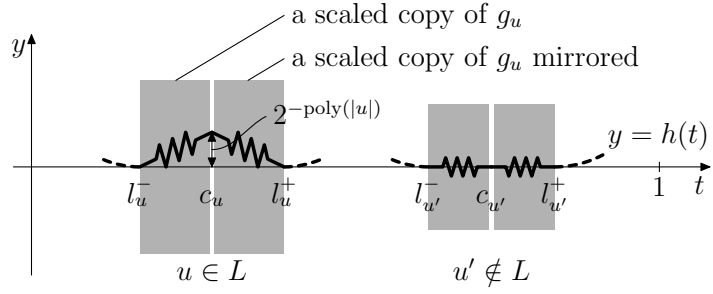}%
 \caption{%
  To construct $g$, 
  we assign interval $[l ^- _u, l ^+ _u]$ to each string~$u$ 
  and put there a pair of reduced copies of $g _u$. 
  The value $h (c _u)$ at the midpoint will be 
  positive if and only if $u \in L$. 
 }
 \label{figure: patchwork}
\end{center}
\end{figure}

  Using this family, 
  the functions $g$ and $h$ in \ref{theorem: main} will be 
  constructed roughly as follows. 
  Divide $[0, 1)$ into 
  infinitely many subintervals $[l ^- _u, l ^+ _u]$, 
  one for each $u$, 
  with midpoints~$c _u$. 
  We put a pair of scaled copies of $g _u$
  onto $[l ^- _u, c _u]$ and $[c _u, l ^+ _u]$
  as shown in \ref{figure: patchwork} 
  so that the membership of $u$ in $L$ can be 
  determined by looking at $h (c _u)$. 
  Scaling down $g _u$ horizontally increases its Lipschitz constant, 
  and the resulting $g$ needs to have a Lipschitz constant independent of $u$; 
  this is why we had to claim in \ref{enumi: Lipschitz} 
  that the $g _u$ originally have 
  small constant $2 ^{-\lambda (\lvert u \rvert)}$. 
  Details are routine 
  and are relegated to \ref{section: proofs}
  along with the proofs of 
  two other theorems which will be stated in \ref{section: variants}
  and which will also follow from \ref{lemma: main}. 

\subsection{Discrete initial value problem and the Lipschitz condition}
 \label{subsection: discrete ivp and the Lipschitz condition}

\begin{figure}
\begin{center}
 \includegraphics[clip,scale=1.05]{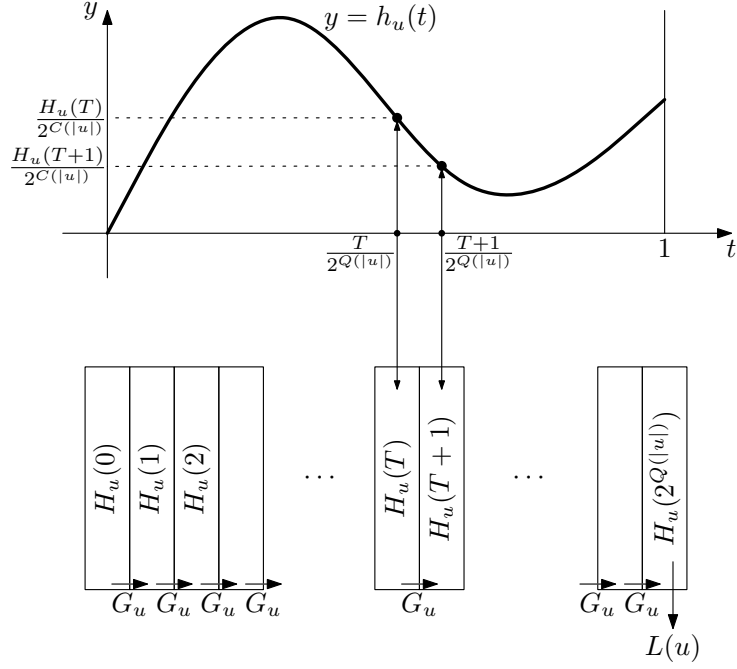}%
 \caption{%
  An attempt to simulate a polynomial-space Turing machine by an initial value problem
  is to encode the machine configuration $H _u (T)$ at each time $T$
  into the value $h _u (t) = H _u (T) / 2 ^{C (\lvert u \rvert)}$
  at time $t = T / 2 ^{Q (\lvert u \rvert)}$. 
 }
 \label{figure: attempt}
\end{center}
\end{figure}

  A first attempt to prove \ref{lemma: main} would be as follows. 
  Consider a polynomial-space 
  Turing machine that decides whether a given string~$u$ belongs to $L$. 
  Its configuration at each time can be encoded into a 
  nonnegative integer less than $2 ^{C (\lvert u \rvert)}$, 
  where $C$ is a polynomial. 
  There is a simple rule that maps 
  $u$ (the input), 
  $T$ (time) and 
  $d$ (the current configuration) to 
  a number $G _u (T, d)$ (the next configuration) such that 
  the recurrence
\begin{equation}
 \label{equation: naive discrete ivp}
  H _u (0) = 0, \qquad
  H _u (T + 1) = G _u \bigl( T, H _u (T) \bigr)
\end{equation}
  leads to $H _u (2 ^{Q (\lvert u \rvert)}) = L (u)$
  for some polynomial $Q$. 
  Now this situation looks similar to 
  the one in \ref{lemma: main}: 
  starting at $0$, the value of $H _u$ (or $h _u$) 
  changes over time according to a simpler function $G _u$ (or $g _u$), 
  to reach a value eventually that indicates the answer $L (u)$. 
  Thus we are tempted to simulate the 
  ``discrete initial value problem''~\ref{equation: naive discrete ivp}
  by embedding each value~$H _u (T)$ 
  as real number~$H _u (T) / 2 ^{C (\lvert u \rvert)}$
  (\ref{figure: attempt}). 

  The obstacle to this attempt is that
  the differential equation 
  \short\ref{enumi: equation} in \ref{lemma: main}
  cannot express all discrete recurrences of form~\ref{equation: naive discrete ivp}: 
  continuous trajectories cannot branch or cross one another; 
  besides, we have the Lipschitz condition~\short\ref{enumi: Lipschitz}
  that puts restriction on how strong 
  the feedback of $h$ to itself can be. 
  We thus need to restrict the discrete problem~\ref{equation: naive discrete ivp} 
  so that it can be simulated by the continuous version. 

  To do so, 
  let us reflect on what the Lipschitz condition~\short\ref{enumi: Lipschitz} means. 
  A rough calculation shows that 
  if two trajectories differ by $\varepsilon$ at time $t$, 
  then 
  they can differ at time $t + 2 ^{-Q (\lvert u \rvert)}$ 
  by at most $
 \varepsilon \exp (2 ^{-\lambda (\lvert u \rvert)} 2 ^{-Q (\lvert u \rvert)})
\approx
 \varepsilon (1 + 2 ^{-\lambda (\lvert u \rvert) - Q (\lvert u \rvert)})
  $.  Thus, 
  the gap can only widen (or narrow) by a factor of 
  $\pm 2 ^{-\lambda (\lvert u \rvert) - Q (\lvert u \rvert)}$ 
  during each time interval of length $2 ^{-Q (\lvert u \rvert)}$. 
  In other words, 
  the feedback caused by equation~\short\ref{enumi: equation} is so weak that 
  each digit of $h _u$ can only affect 
  far lower digits of $h _u$ in the next step. 

\begin{figure}
\begin{center}
 \includegraphics[clip,scale=1.10]{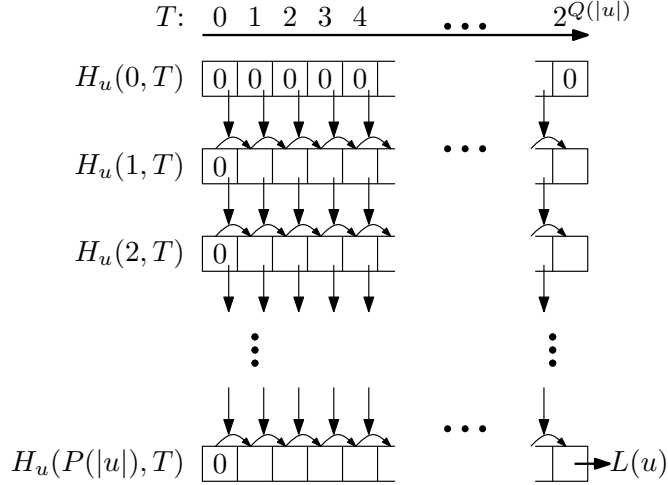}%
 \caption{%
  The discrete initial value problem 
  \ref{equation: discrete ivp type 1}--\ref{equation: discrete ivp main}. 
  Each cell $H _u (T)$ in \ref{figure: attempt} 
  is now divided into $H _u (0, T)$, \ldots, $H _u (P (\lvert u \rvert), T)$; 
  the increment from $H _u (i + 1, T)$ to $H _u (i + 1, T + 1)$
  is computed by $G _u$ using the upper left cell $H _u (i, T)$. 
 }
 \label{figure: discrete_ivp}
\end{center}
\end{figure}

  Now we define a discrete problem 
  that reflects this restriction. 
  Let $P$ and $Q$ be polynomials and let
\begin{align}
 \label{equation: discrete ivp type 1}
 G _u 
& 
 \tcolon 
  [P (\lvert u \rvert)] \times [2 ^{Q (\lvert u \rvert)}] \times [4]
 \tto
  \{-1, 0, 1\}, 
\\
 \label{equation: discrete ivp type 2}
 H _u 
& 
 \tcolon 
  [P (\lvert u \rvert) + 1] \times [2 ^{Q (\lvert u \rvert)} + 1] 
 \tto
  [4], 
\end{align}
  where we write $[N] = \{0, \dots, N - 1\}$ for $N \in \Nset$. 
  Our restricted discrete initial value problem is as follows: 
\begin{align}
 \label{equation: discrete ivp init}
  H _u (i, 0) 
&
 =
  H _u (0, T)
 =
  0, 
\\
 \label{equation: discrete ivp main}
  H _u (i + 1, T + 1)
&
 =
  H _u (i + 1, T) + G _u \bigl( i, T, H _u (i, T) \bigr). 
\end{align}
  Thus, $H _u (T)$ of \ref{equation: naive discrete ivp}
  is now divided into polynomially many (in $\lvert u \rvert$) components 
  $H _u (0, T)$, \ldots, $H _u (P (\lvert u \rvert), T)$; 
  compare Figures \bare\ref{figure: attempt} (bottom) 
  and \bare\ref{figure: discrete_ivp}. 
  We have added the restriction that 
  $G _u$ sees only the component $H _u (i, T)$, 
  which in \ref{figure: discrete_ivp} means the upper left of the current cell. 
  The following lemma states that, despite this restriction, 
  we still have $\classPSPACE$-completeness. 
  Note that making $G _u$ completely oblivious to 
  its last argument would be an overkill, 
  because then $H _u$ would just add up the values of $G _u$, 
  resulting in the complexity merely of $\classNumberP$. 

\begin{lemma}
 \label{lemma: discrete}
  Let $L \in \classPSPACE$. 
  Then there are polynomials $P$, $Q$ and 
  families $(G _u) _u$, $(H _u) _u$ satisfying 
  \ref{equation: discrete ivp type 1}--\ref{equation: discrete ivp main} 
  such that $(G _u) _u$ is polynomial-time computable 
  and $
H _u (P (\lvert u \rvert), \allowbreak 2 ^{Q (\lvert u \rvert)}) = L (u) % ad hoc
  $ for each string~$u$. 
\end{lemma}

  Before proving this, 
  we will reduce \ref{lemma: main} to \ref{lemma: discrete}
  by simulating the new system 
  \ref{equation: discrete ivp type 1}--\ref{equation: discrete ivp main} 
  by the differential equation. 
  Using $G _u$ and $H _u$ of \ref{lemma: discrete}, 
  we will construct $g _u$ and $h _u$ of \ref{lemma: main}
  such that 
  $h _u (T / 2 ^{Q (\lvert u \rvert)}) = \sum _i H _u (i, T) / B ^i$
  for each $T$, 
  where $B$ is a big number. 
  Thus, each column in \ref{figure: discrete_ivp} 
  will be encoded into one real number 
  so that upper/lower cells in the column correspond to 
  upper/lower bits of the real number. 
  Thanks to the restriction that $G _u$ sees only the upper row, 
  the differential equation $h' _u (t) = g _u (t, h _u (t))$ only needs to cause 
  a weak feedback on $h _u$ 
  where each bit of the value of $h _u$ 
  affects only much lower bits of its next value. 
  This keeps $g _u$ Lipschitz continuous. 
  Now we fill in the details. 

\begin{namedproof}{Proof of \ref{lemma: main}}
  Let $P$, $Q$, $(G _u) _u$, $(H _u) _u$ be
  as in \ref{lemma: discrete}. 
  By ``dividing each unit time into $P (\lvert u \rvert)$ steps,''
  we may assume that for each $T$, there is at most one $i$ 
  such that $G _u (i, T, Y) \neq 0$ for some $Y$. 
  Write $j _u (T)$ for this unique~$i$
  (define $j _u (T)$ arbitrarily 
  if there is no such $i$). 
  We may further assume that 
\begin{equation}
 \label{equation: discrete ivp final value}
 H _u \bigl( i, 2 ^{Q (\lvert u \rvert)} \bigr) = 
\begin{cases}
 L (u) & \text{if} \ i = P (\lvert u \rvert), \\
 0     & \text{if} \ i < P (\lvert u \rvert). 
\end{cases}
\end{equation}
  Thus, not only does the bottom right corner of \ref{figure: discrete_ivp} 
  equal $L (u)$, 
  as stated already in \ref{lemma: discrete}, 
  but we also claim that the other cells in the rightmost column are all $0$. 
  This can be achieved by 
  doubling the time frame and 
  extending $G$ symmetrically 
  so that in the second half it cancels out what it has done. 
  Precisely, 
  we extend $G _u$ by 
\begin{equation}
 G _u (i, 2 \cdot 2 ^{Q (\lvert u \rvert)} - 1 - T, Y)
=
\begin{cases}
 0               & \text{if} \ i = P (\lvert u \rvert) - 1, \\
 -G _u (i, T, Y) & \text{if} \ i < P (\lvert u \rvert) - 1
\end{cases} 
\end{equation}
  for $
(i, T, Y) \in 
 [P (\lvert u \rvert)] \times [2 ^{Q (\lvert u \rvert)}] \times [4]
  $, 
  and 
  $H _u$ by
\begin{equation}
 H _u (i, 2 \cdot 2 ^{Q (\lvert u \rvert)} - T)
=
\begin{cases}
 H _u \bigl( P (\lvert u \rvert), 2 ^{Q (\lvert u \rvert)} \bigr) & \text{if} \ i = P (\lvert u \rvert), \\
 H _u (i, T) & \text{if} \ i < P (\lvert u \rvert) 
\end{cases}
\end{equation}
  for $(i, T) \in [P (\lvert u \rvert) + 1] \times [2 ^{Q (\lvert u \rvert)} + 1]$, 
  and then add $1$ to $Q (\lvert u \rvert)$. 
  It is easy to verify that the equations
  \ref{equation: discrete ivp init} and \ref{equation: discrete ivp main}
  are still satisfied. 

  Now, assuming \ref{equation: discrete ivp final value}, 
  we construct the families 
  $(g _u) _u$ and $(h _u) _u$ of \ref{lemma: main}. 
  For each string $u$ and each $(t, y) \in [0, 1] \times [-1, 1]$, 
  let 
  $T \in \Nset$, 
  $\theta \in [0, 1]$, 
  $Y \in \Zset$, 
  $\eta \in [-1 / 4, 3 / 4]$ be such that 
  $t = (T + \theta) 2 ^{-Q (\lvert u \rvert)}$ and 
  $y = (Y + \eta) B ^{-j _u (T)}$, 
  and define 
\begin{align}
 \label{equation: 0802100213}
  g _u (t, y)
&
 =
\begin{cases}
  \displaystyle
   \frac{2 ^{Q (\lvert u \rvert)} \pi \sin (\theta \pi)}{2 B ^{j _u (T) + 1}} 
   G _u \bigl( j _u (T), T, Y \bmod 4 \bigr) 
& 
  \displaystyle
  \text{if} \ 
  \eta \leq \frac 1 4, 
\\[9pt]
  \displaystyle
     \frac{3 - 4 \eta}{2}
     g _u \biggl( 
      t, 
      \frac{Y}{B ^{j _u (T)}} 
     \biggr)
   +
     \frac{4 \eta - 1}{2}
     g _u \biggl( 
      t, 
      \frac{Y + 1}{B ^{j _u (T)}} 
     \biggr) 
&
  \displaystyle
  \text{if} \ 
  \eta \geq \frac 1 4, 
\end{cases}
\\
 \label{equation: 0802100220}
  h _u (t)
&
 =
    \frac{1 - \cos (\theta \pi)}{2} 
   \cdot
    \frac{G _u \bigl( j _u (T), T, H _u (j _u (T), T) \bigr)}
         {B ^{j _u (T) + 1}}
  +
   \sum _{i = 0} ^{P (\lvert u \rvert)} \frac{H _u (i, T)}{B ^i}, 
\end{align}
  where $B = 2 ^{\lambda (\lvert u \rvert) + Q (\lvert u \rvert) + 5}$. 
  Note that the second branch of \ref{equation: 0802100213} says that 
  when $\eta \in [1 / 4, 3 / 4]$, 
  we define $g _u (t, y)$ by interpolating between 
  the nearest two $y$ at which $g _u$ is already defined by the first branch. 
  Equation \ref{equation: 0802100220} says that, 
  when $\theta = 0$ (i.e., $t$ is a multiple of $2 ^{-Q (\lvert u \rvert)}$), 
  the value $h _u (t)$ is the real number that encodes
  the $T$th column of \ref{figure: discrete_ivp}; 
  as $\theta$ goes from $0$ to $1$, 
  it moves to the next value along a cosine curve, 
  whose slope, as we will see below, 
  matches the sine function in the first branch of \ref{equation: 0802100213}. 
  It is easy to verify that
  the definition is consistent; 
  in particular, 
  we use \ref{equation: discrete ivp main} to show that 
  \ref{equation: 0802100220} stays the same 
  for the two choices of $(T, \theta)$ 
  when $t$ is a multiple of $2 ^{-Q (\lvert u \rvert)}$. 

  Conditions \short\ref{enumi: range} and \short\ref{enumi: boundary} 
  of \ref{lemma: main} are easy to verify. 
  We have \short\ref{enumi: output} with $
\rho (k) = P (k) (\lambda (k) + Q (k) + 5)
  $, 
  since $
 h _u (1)
=
 H _u (P (\lvert u \rvert), 2 ^{Q (\lvert u \rvert)}) / B ^{P (\lvert u \rvert)}
=
 L (u) / B ^{P (\lvert u \rvert)}
=
 L (u) / 2 ^{\rho (\lvert u \rvert)}
  $ by \ref{equation: 0802100220} and \ref{equation: discrete ivp final value}. 
  Checking the polynomial-time computability of $(g _u) _u$ is also routine, 
  using \ref{lemma: polynomial modulus}. 

\pagebreak[1] % ad hoc 

  To see \short\ref{enumi: equation}, 
  observe that 
  in the right-hand side of \ref{equation: 0802100220}, 
\begin{itemize}
 \item 
  the absolute value of the first term is bounded by $
B ^{-j _u (T) - 1} \leq B ^{-j _u (T)} / 32
  $, 
 \item 
  the summands corresponding to $i \leq j _u (T)$ are
  multiples of $
B ^{-j _u (T)}
  $, and 
 \item 
  the summands corresponding to $i > j _u (T)$ are nonnegative numbers, 
  each bounded by $
 3 / B ^i 
=
 3 B ^{-j _u (T)} / B ^{i - j _u (T)}
\leq 
 3 B ^{-j _u (T)} / 32 ^{i - j _u (T)}
  $, and thus altogether by $3 B ^{-j _u (T)} / 31$. 
\end{itemize}
  Hence, we can write $
h _u (t) = (Y + \eta) B ^{-j _u (T)} 
  $ for some $\eta \in [-1 / 4, 1 / 4]$, where 
\begin{equation}
  Y = \sum _{i = 0} ^{j _u (T)} H _u (i, T) \cdot B ^{j _u (T) - i}. 
\end{equation}
  Since $B$ is a multiple of $4$, 
  we have $Y \bmod 4 = H _u (j _u (T), T)$. 
  Substituting these $Y$ and $\eta$ into (the first branch of) 
  \ref{equation: 0802100213}, we get 
\begin{equation}
  g _u \bigl( t, h _u (t) \bigr)
 =
  \frac{2 ^{Q (\lvert u \rvert)} \pi \sin (\theta \pi)}{2 B ^{j _u (T) + 1}} 
  G _u \bigl( j _u (T), T, H _u (j _u (T), T) \bigr). 
\end{equation}
  This equals $h' _u (t)$ calculated from \ref{equation: 0802100220}. 

  For the Lipschitz condition \short\ref{enumi: Lipschitz}, 
  note that 
  since the value of $G _u$ in the first branch of \ref{equation: 0802100213} 
  is in $\{-1, 0, 1\}$, 
  the difference between the two values of $g _u$ 
  in the second branch 
  is bounded by $
 2 \times 2 ^{Q (\lvert u \rvert)} \pi \sin (\theta \pi) / (2 B ^{j _u (T) + 1})
<
 2 ^{Q (\lvert u \rvert) + 2} / B ^{j _u (T) + 1}
  $.  
  Thus, the slope of $g _u$ along the second argument is at most 
\begin{equation}
 2 B ^{j _u (T)} \cdot \frac{2 ^{Q (\lvert u \rvert) + 2}}{B ^{j _u (T) + 1}}
=
 \frac{2 ^{Q (\lvert u \rvert) + 3}}{B}
% =
%  2 ^{-C (\lvert u \rvert) - \lambda (\lvert u \rvert)}
\leq
 2 ^{-\lambda (\lvert u \rvert)}
\end{equation}
  by our choice of $B$. 
\end{namedproof}

\subsection{The discrete initial value problem is hard}
 \label{subsection: discrete ivp is hard}

  It remains to prove \ref{lemma: discrete}. 
  At first sight, 
  our system 
  \ref{equation: discrete ivp type 1}--\ref{equation: discrete ivp main}
  (\ref{figure: discrete_ivp})
  looks too weak to simulate 
  a polynomial-space computation: 
  although we have polynomial amount of memory (rows)
  and exponential amount of time (columns), 
  the ``chains of dependence'' of values 
  must run from top to bottom
  and thus are polynomially bounded in length. 

  Thus, we give up embedding a general $\classPSPACE$ computation
  into \ref{figure: discrete_ivp}. 
  Instead, we embed another $\classPSPACE$-complete problem, 
  $\probQBF$ (quantified boolean formulas, 
  see
\cite{papadimitriou94:_comput_compl}
  where it is called \textsc{qsat}), 
  which asks for 
  the truth value of the given formula $u$ of form
\begin{equation}
\label{equation: qbf}
Q _n x _n \dots Q _1 x _1 \ldotp \psi (x _1, \dots, x _n), 
\end{equation}
  where $\psi$ is a boolean formula and 
  $Q _i \in \{\forall, \exists\}$ 
  for each $i = 1$, \ldots, $n$. 

  The truth value of \ref{equation: qbf} is obtained by
  evaluating a binary tree of depth $n$ 
  whose $2 ^n$ leaves each correspond to an assignment to $(x _1, \ldots, x _n)$ and 
  whose internal nodes at level~$i$ are labeled $Q _i$. 
  This is roughly why it can be simulated by 
  the tableau in \ref{figure: discrete_ivp} 
  despite the restriction that the dependence of values 
  must run from top to bottom. 
  We give a formal proof and then an example (\ref{figure: qbf_encoding_example}). 

\begin{namedproof}{Proof of \ref{lemma: discrete}}
  We may assume that $L = \probQBF$. 
  We will construct the polynomials $P$, $Q$ and 
  families $(G _u) _u$ and $(H _u) _u$ in \ref{lemma: discrete}. 
  Let $u$ be of form \ref{equation: qbf}. 
  For each $i = 0$, \ldots, $n$ 
  and $n - i$ bits $b _{i + 1}$, \ldots, $b _n \in \{0, 1\}$, 
  we write $\psi _i (b _{i + 1}, \dots, b _n) \in \{0, 1\}$ for 
  the truth value ($1$ for true) of the subformula $
Q _i x _i \dots Q _1 x _1 \ldotp \allowbreak \psi (x _1, \dots, x _i, b _{i + 1}, \ldots, b _n)
  $, 
  so that $\psi _0 = \psi$ and $\psi _n (\,) = \probQBF (u)$. 
  We regard quantifiers as functions from $\{0, 1, 2, 3\}$ to $\{0, 1\}$: 
\begin{align}
 \label{equation: meaning of quantifiers}
&
  \forall (2) = \exists (2) = \exists (1) = 1, 
&
&
  \exists (0) = \forall (0) = \forall (1) = 0 
\end{align}
  (the values $\forall (3)$ and $\exists (3)$ do not matter). 
  These correspond to the meaning of the quantifiers: 
  $\forall x$ (resp.\ $\exists x$) means that the subsequent formula is 
  satisfied by $2$ (resp.\ $1$ or $2$) of the two possible assignments to $x$. 
  Thus, 
\begin{equation}
 \label{equation: quantifier step}
  Q _{i + 1} \bigl( 
    \psi _i (0, b _{i + 2}, \ldots, b _n) 
   +
    \psi _i (1, b _{i + 2}, \ldots, b _n) 
  \bigr) 
 =
  \psi _{i + 1} (b _{i + 2}, \ldots, b _n)
\end{equation}
  for each $i = 0$, \ldots, $n - 1$. 
  For $2 n + 1$ bits $b _0$, \ldots, $b _{2 n} \in \{0, 1\}$, 
  we write $\overline{b _{2 n} \ldots b _0}$ 
  for the number $b _0 + 2 b _1 + 2 ^2 b _2 + \dots + 2 ^{2 n} b _{2 n}$. 

  To define $G _u$, 
  let 
\begin{multline}
 \label{equation: 0802100302}
  G _u (i, \overline{T _{2 n} T _{2 n - 1} \ldots T _{2 i + 2} T _{2 i + 1} 1 0 0 \ldots 0}, Y) 
\\
 = 
  (-1) ^{T _{2 i + 2}} \times 
\begin{cases}
  \psi _0 (T _1 \oplus T _2, T _3 \oplus T _4, \dots, T _{2 n - 1} \oplus T _{2 n})
&
  \text{if} \ i = 0, 
\\
  Q _i (Y) 
  \qquad\qquad\qquad\qquad
&
  \text{otherwise}, 
\end{cases}
\end{multline}
  where $\mathord\oplus$ denotes the exclusive or; 
  let $G _u (i, T, Y) = 0$ for other $T$ 
  (that is, when $T$ is not an odd multiple of $2 ^{2 i}$). 
  Define $H _u$ from $G _u$ by 
  \ref{equation: discrete ivp init} and \ref{equation: discrete ivp main}. 

  We prove by induction on $i = 0$, \ldots, $n$ that 
  $H _u (i, T) \in \{0, 1, 2\}$ for all $T$, as we mentioned earlier, 
  and that 
\begin{equation}
 \label{equation: induction hypothesis}
 G _u (i, S, H _u (i, S)) 
=
  (-1) ^{S _{2 i + 2}} 
  \psi _i (S _{2 i + 1} \oplus S _{2 i + 2}, \dots, S _{2 n - 1} \oplus S _{2 n}) 
\end{equation}
  for all $S$ of form 
  $\overline{S _{2 n} S _{2 n - 1} \ldots S _{2 i + 1} 1 0 0 \ldots 0}$ 
  (it is immediate from the definition of $G _u$ that 
  $G _u (i, S, H _u (i, S)) = 0$ for other $S$). 
  Once we have proved this, 
  the case $i = n$ yields 
  $G _u (n, 2 ^{2 n}, H _u (n, 2 ^{2 n})) = \psi _n (\,) = \probQBF (u)$, 
  and hence $H _u (n + 1, 2 ^{2 n} + 1) = \probQBF (u)$. 
  Since $n < \lvert u \rvert$, 
  we can add dummy rows and columns 
  so that $
H _u (P (\lvert u \rvert), 2 ^{Q (\lvert u \rvert)}) = \probQBF (u)
  $ for some polynomials $P$ and $Q$, as required. 

  The claims for $i = 0$ follow immediately from 
  \ref{equation: discrete ivp init} and \ref{equation: 0802100302}. 
  Now suppose \ref{equation: induction hypothesis} as the induction hypothesis 
  and fix $T = \overline{T _{2 n} T _{2 n - 1} \ldots T _0}$. 
  Let $
 Y = H _u (i + 1, T)
  $. 
  By \ref{equation: discrete ivp init} and \ref{equation: discrete ivp main}, 
  we have
\begin{equation}
 \label{equation: carry}
  Y
 =
  \sum _{S = 0} ^{T - 1} G _u \bigl( i, S, H _u (i, S) \bigr). 
\end{equation}
  Since the assumption \ref{equation: induction hypothesis} implies that 
  flipping the two bits $S _{2 i + 2}$ and $S _{2 i + 1}$
  of any $S = \overline{S _{2 n} S _{2 n - 1} \ldots S _0}$ 
  reverses the sign of $G _u (i, S, H _u (i, S))$, 
  most of the nonzero summands in \ref{equation: carry} cancel out. 
  The only terms that can survive are 
  those 
  that correspond to 
  $S = \overline{T _{2 n} T _{2 n - 1} \ldots T _{2 i + 3} 0 0 1 0 0 \ldots 0}$ and 
  $S = \overline{T _{2 n} T _{2 n - 1} \ldots T _{2 i + 3} 0 1 1 0 0 \ldots 0}$. 
  This proves $Y \in \{0, 1, 2\}$. 

  When 
  $T = \overline{T _{2 n} T _{2 n - 1} \ldots T _{2 i + 3} 1 0 0 \ldots 0}$, 
  both of these terms survive, 
  so that
\begin{equation}
  Y
 =
   \psi _i (0, T _{2 i + 3} \oplus T _{2 i + 4}, \dots, T _{2 n - 1} \oplus T _{2 n}) 
  + 
   \psi _i (1, T _{2 i + 3} \oplus T _{2 i + 4}, \dots, T _{2 n - 1} \oplus T _{2 n}). 
\end{equation}
  Therefore, $
  Q _{i + 1} (Y)
 =
   \psi _{i + 1} (T _{2 i + 3} \oplus T _{2 i + 4}, \dots, T _{2 n - 1} \oplus T _{2 n}) 
  $ by \ref{equation: quantifier step}. 
  Thus, 
\begin{equation}
 \label{equation: induction goal}
 G _u (i + 1, T, Y) 
=
  (-1) ^{T _{2 i + 4}} 
   \psi _{i + 1} (T _{2 i + 3} \oplus T _{2 i + 4}, \dots, T _{2 n - 1} \oplus T _{2 n}) 
\end{equation}
  by \ref{equation: 0802100302}, 
  completing the induction step. 
\end{namedproof}

\begin{figure}
\begin{center}
 \includegraphics[clip,width=\textwidth]{discrete_ivp_cm.eps}%
 \caption{%
  The discrete initial value problem 
  \ref{equation: discrete ivp type 1}--\ref{equation: discrete ivp main}
  corresponding to
  the formula $u = \exists x _2 \ldotp \forall x _1 \ldotp (x _1 \lor x _2)$.
  We follow the convention in \ref{figure: discrete_ivp}
  (but omit the top cells $H _u (0, T)$ which are always $0$): 
  the cells contain $H _u (i, T)$, and 
  the signed number above the cell $H _u (i + 1, T)$ indicates 
  the increment $G _u (i, T, H _u (i, T))$ (which is $0$ when omitted). 
  The increments $G _u (0, T, 0)$ for the top row are determined by 
  the truth values of $x _1 \lor x _2$ for various assignments to $(x _1, x _2)$. 
 }
 \label{figure: qbf_encoding_example}
\end{center}
\end{figure}

  \ref{figure: qbf_encoding_example} shows the table 
  when $u$ be the formula
  $\exists x _2 \ldotp \forall x _1 \ldotp (x _1 \lor x _2)$. 
  The values $G _u (0, T, 0)$
  encode (redundantly) 
  the truth table of the matrix $x _1 \lor x _2$ 
  (first branch of \ref{equation: 0802100302}). 
  For example, $G _u (0, T, 0) = \pm 1$ (resp.\ $0$) 
  for $T = 3, 5, 27, 29$ (resp.\ $1, 7, 25, 31$)
  because $(x _1, x _2) = (1, 0)$ (resp.\ $(0, 0)$) 
  makes $x _1 \lor x _2$ true (resp.\ false). 
  Also observe that $H _u (1, T)$ returns to $0$ every eight cells. 
  As a result, 
  the cell $H _u (1, 4) = 1$ (resp.\ $H _u (1, 12) = 2$) represents the fact that 
  when $x _2$ is false (resp.\ true), 
  $x _1 \lor x _2$ is satisfied by 
  one (resp.\ two) of the assignments to $x _1$. 
  Now look at the next row. 
  The second branch of \ref{equation: 0802100302} says that 
  for odd multiples $T$ of $4$, 
  the values $G _u (1, T, H _u (1, T))$ 
  are $\pm 1$ or $0$
  according to whether the upper left cell has a $2$ or not. 
  Thus, they encode the smaller truth table for 
  the subformula $\forall x _1 \ldotp (x _1 \lor x _2)$ 
  under each assignment to $x _2$. 
  As a result, 
  the cell $H _u (2, 16) = 1$ indicates that 
  this subformula is satisfied by one of the assignments to $x _2$, 
  which causes the last row to get incremented at $T = 17$.
  Observe that the final cell $H _u (3, 32)$ has a $1$, 
  exactly because $u$ is true. 

\section{Other versions of the problem}
 \label{section: variants}

\subsection{Complexity of the final value}

\cite{ko83:_comput_compl_of_ordin_differ_equat}
  discusses another version of the question
  which relates the complexity of 
  the value~$h (1)$, rather than the function~$h$, 
  to that of tally languages (subsets of $\{0\} ^*$). 

  \ref{definition: completeness} extends straightforwardly to 
  machines taking $d$ oracles. 
  In particular, the case $d = 0$ means that 
  a language~$L$ is said to reduce to a real number~$t$ if 
  there are polynomial-time computable functions $R$ and $S$
  such that 
  $L (u) = R (A (S (u)))$ for any string~$u$ and any name~$A$ of $t$. 
  Now we can state: 

\begin{theorem}
 \label{theorem: main 2}
  For any tally language $L \in \classPSPACE$, 
  there are functions $g$ and $h$ satisfying \kome{}
  such that $L$ reduces to $h (1)$. 
\end{theorem}

  This can be proved by
  arranging the building blocks $g _u$ from \ref{lemma: main}, 
  as we did for \ref{theorem: main}, 
  but in a different way. 
  See \ref{subsection: proof of theorem main 2} for details. 

  As a corollary to \ref{theorem: main 2}, 
  all tally languages of $\classPSPACE$ are in $\classP$ if 
  \kome{} implies that $h (1)$ is polynomial-time computable. 
  This improves on Theorem~4 of 
\cite{ko83:_comput_compl_of_ordin_differ_equat}, 
  which showed the same thing with 
  the Lipschitz condition replaced by a weaker condition. 

\subsection{Volterra integral equations}
 \label{subsection: Volterra}
\cite{ko92:_comput_compl_of_integ_equat}
  later studied
  the complexity of \emph{Volterra integral equations of the second kind}
\begin{equation}
 \label{equation: Volterra second}
 h (t) = f (t) + \int _0 ^t g \bigl( t, \tau, h (\tau) \bigr) \, \d \tau, 
\qquad
 t \in [0, 1], 
\end{equation}
  where function $h$ is to be solved from 
  given $f$ and $g$. 
  As before, 
  we suppose that $f$ and $g$ are polynomial-time computable 
  and ask how complex $h$ can be. 

  If $g$ is Lipschitz continuous (in its last argument), 
  $h$ is polynomial-space computable 
  by Picard's iteration method (Section~3 of 
\cite{ko92:_comput_compl_of_integ_equat}). 
  On the other hand, 
  the best lower bound (in the sense 
  analogous to the ``if'' direction of \ref{corollary: answer to Ko}) 
  has been $\classP ^{\classNumberP}$. 
  One of the two open problems in 
\cite{ko92:_comput_compl_of_integ_equat}
%  (posed in the last paragraph of Section~1 there)
  was to close this gap. 
  Our \ref{theorem: main} has solved it, 
  because the initial value problem~\ref{equation: problem}
  is the special case of \ref{equation: Volterra second} 
  where $f$ is constantly zero and $g$ ignores its first argument: 

\begin{corollary}
  There are functions
  $f \tcolon [0, 1] \tto \Rset$ and 
  $g \tcolon [0, 1] \times [0, 1] \times \Rset \tto \Rset$ such that 
  $g$ is Lipschitz continuous (in its last argument), 
  $f$ and $g$ are both polynomial-time computable, 
  and the (unique) solution of \ref{equation: Volterra second} is 
  polynomial-space complete. 
\end{corollary}

  The other problem was about 
  the following weak version of the Lipschitz condition~\ref{equation: Lipschitz}: 
\begin{equation}
 \label{equation: local Lipschitz}
% \tag{\bare\ref{equation: Lipschitz}$'$}
 \lvert g (t, y _0) - g (t, y _1) \rvert \leq 2 ^{r (n)} \lvert y _0 - y _1 \rvert, 
\quad
 n \in \Nset, \ 
 t \in [0, 1 - 2 ^{-n}], \ 
 y _0, y _1 \in \Rset, 
\end{equation}
  where $r$ is a polynomial. 
  Assuming \kome{} with \ref{equation: Lipschitz} 
  replaced by \ref{equation: local Lipschitz}, 
  how complex can $h$ be, 
  provided it has a polynomial modulus of continuity? 
\cite{ko92:_comput_compl_of_integ_equat}
  asked this question for 
  the Volterra equation~\ref{equation: Volterra second}
  (in which case the $g$ in \ref{equation: local Lipschitz} 
  takes one more argument), 
  and showed that $h$ is exponential-space computable 
  and can be $\classEXP$-hard. 
  His second open problem was to close this gap. 

  The motivation for this problem comes from 
\emph{Volterra integral equations of the first kind}, 
  a class of equations that are considered harder to solve
  than~\ref{equation: Volterra second}. 
  A common approach to solve them is to convert the equation 
  into the form~\ref{equation: Volterra second}. 
  This conversion does not preserve 
  Lipschitz continuity~\ref{equation: Lipschitz}; 
  the new equation merely satisfies \ref{equation: local Lipschitz}. 
  See 
\cite{ko92:_comput_compl_of_integ_equat}
  for details. 

The following theorem settles this problem
(see \ref{subsection: proof of theorem main exp} for a proof, 
again using \ref{lemma: main}). 
In fact, we have $\classEXPSPACE$-completeness
even for the simple differential equation~\ref{equation: problem}: 

\begin{theorem}
 \label{theorem: main exp}
  There are functions $g$ and $h$ satisfying~\kome{} 
  with \ref{equation: Lipschitz} replaced with \ref{equation: local Lipschitz}
  such that $h$ has a polynomial modulus of continuity 
  and yet is $\classEXPSPACE$-hard. 
\end{theorem}

  Note, however, that this theorem says nothing directly about
  the complexity of Volterra equations of the first kind. 
  It merely addresses the complexity of a class of equations 
  that may arise in a particular approach to solving them. 

\section{Putting the building blocks together}
 \label{section: proofs}

We give the details of the proofs of Theorems 
\bare\ref{theorem: main}, \bare\ref{theorem: main 2} and \bare\ref{theorem: main exp}
from \ref{lemma: main}. 

\subsection{Proof of \ref{theorem: main}}
 \label{subsection: proof of theorem main}
  The rough ideas are explained in 
  \ref{subsection: building blocks}. 
  Let $L$ be a $\classPSPACE$-complete language. 
  Use \ref{lemma: main} for $\lambda (k) = 2 k + 2$
  to obtain polynomial~$\rho$ and families $(g _u) _u$, $(h _u) _u$. 
  Since $(g _u) _u$ is polynomial-time computable, 
  there is a polynomial $\gamma$ 
  satisfying $
\lvert g _u (t, y) \rvert \leq 2 ^{\gamma (\lvert u \rvert) - \lvert u \rvert}
  $.  
  For each binary string~$u$, 
  let $\varLambda _u = 2 ^{\lambda (\lvert u \rvert)}$, 
  $\varGamma _u = 2 ^{\gamma (\lvert u \rvert)}$ and 
\begin{align}
  c _u 
&
 =
   1
  -
   \frac{1}{2 ^{\lvert u \rvert}} 
  + 
    \frac{2 \overline u + 1}{\varLambda _u}, 
&
  l ^\mp _u
&
 = 
  c _u \mp \frac{1}{\varLambda _u}, 
\end{align}
  where $\overline u \in \{0, \dots, 2 ^{\lvert u \rvert} - 1\}$ is 
  $u$ read as an integer in binary notation. 
  This divides $[0, 1)$ into intervals $[l ^- _u, l ^+ _u]$
  indexed by $u \in \{0, 1\} ^*$. 
  Define 
\begin{align}
 \label{equation: vector}
  g \biggl( 
   l ^\mp _u \pm \frac{t}{\varLambda _u},
   \frac{y}{\varLambda _u \varGamma _u} 
  \biggr)
&
 = 
  \pm \frac{g _u (t, \Hat y)}{\varGamma _u}, 
\\
 \label{equation: flow}
  h \biggl( 
   l ^\mp _u \pm \frac{t}{\varLambda _u}
  \biggr)
&
 = 
  \frac{h _u (t)}{\varLambda _u \varGamma _u} 
\end{align}
  for each $t \in [0, 1]$ and $y \in \Rset$, 
  where $\Hat y = \max \{-1, \allowbreak \min\{1, y\}\}$. 
  Let $g (1, y) = h (1) = 0$ for each $y \in \Rset$. 
  These define $g$ and $h$ ``seamlessly'' 
  by \ref{enumi: boundary}. 

  We show that $g$ and $h$ satisfy \kome{}. 
  We begin with equation~\ref{equation: problem}. 
  It is easy to see that $h (0) = 0$ and $h' (1) = 0 = g (1, h (1))$. 
  Since any number in $[0, 1)$ can be written in the form 
  $l ^\mp _u \pm t / \varLambda _u$ for some $u$ and $t \in [0, 1]$, 
  we have \ref{equation: problem} by 
\begin{align}
  h' \biggl( 
   l ^\mp _u \pm \frac{t}{\varLambda _u}
  \biggr)
&
 =
  \pm \frac{h' _u (t)}{\varGamma _u}
 =
  \pm \frac{g _u (t, h _u (t))}{\varGamma _u}
\\
\notag
&
 =
  g \biggl( 
   l ^\mp _u \pm \frac{t}{\varLambda _u},
   \frac{h _u (t)}{\varLambda _u \varGamma _u} 
  \biggr)
 =
  g \biggl( l ^\mp _u \pm \frac{t}{\varLambda _u}, h \biggl( l ^\mp _u \pm \frac{t}{\varLambda _u} \biggr) \biggr), 
\end{align}
  where equalities are by 
  \ref{equation: flow}, 
  \short\ref{enumi: equation}, 
  \ref{equation: vector}, 
  \ref{equation: flow}, 
  respectively. 

  The Lipschitz condition~\ref{equation: Lipschitz} is satisfied
  with $Z = 1$ by \short\ref{enumi: Lipschitz} and our choice of $\lambda$. 
  To see that $g$ is polynomial-time computable, 
  we use 
  (the obvious two-dimensional version of) 
  \ref{lemma: polynomial modulus}. 
  When asked for a $2 ^{-m}$-approximation of $g (T, Y)$ 
  for rational numbers $T$ and $Y$, 
  the machine can 
  find $u$, $\pm$, $t$, $y$ with $
 (T, Y) 
=
 ( 
  l ^\mp _u \pm t / \varLambda _u, \allowbreak
  y / \varLambda _u \varGamma _u 
 )
  $ in polynomial time. 
  Since \ref{equation: vector}
  lies in $[-2 ^{-\lvert u \rvert}, 2 ^{-\lvert u \rvert}]$, 
  the machine can safely answer $0$ if $m < \lvert u \rvert$. 
  Otherwise it can answer by 
  computing $g _u (t, \Hat y)$ with precision $2 ^{-m}$, 
  which can be done in time polynomial in $
m + \lvert u \rvert \leq 2 m
  $ by the polynomial-time computability of $(g _u) _u$. 

  We have thus proved \kome{}. 
  Since
\begin{equation}
 \label{equation: output}
 h (c _u) 
=
 \frac{h _u (1)}{\varLambda _u \varGamma _u}
=
 \frac{L (u)}{2 ^{\lambda (\lvert u \rvert) + \gamma (\lvert u \rvert) + \rho (\lvert u \rvert)}} 
\end{equation}
  by \ref{equation: flow} and \short\ref{enumi: output}, 
  the problem~$L$ reduces to $h$. 
  More precisely, 
  the functions $R$, $S$, $T$ in \ref{definition: completeness}
  can be given by
\begin{align}
  R (u, v) 
&
 = 
\begin{cases}
 0 & \text{if} \ v \ \text{denotes} \ 0, \\
 1 & \text{if} \ v \ \text{denotes} \ 1, 
\end{cases}
\\
  S (u, 0 ^n)
&
 = 
  \text{a string denoting} \ \lfloor 2 ^n c _u \rfloor, 
\\
  T (u)
&
 = 
  0 ^{\lambda (\lvert u \rvert) + \gamma (\lvert u \rvert) + \rho (\lvert u \rvert)}. 
\end{align}
  Since $L$ is $\classPSPACE$-complete, 
  so is $h$. 

\subsection{Proof of \ref{theorem: main 2}}
 \label{subsection: proof of theorem main 2}

\begin{figure}
\begin{center}
 \includegraphics[clip,scale=.91]{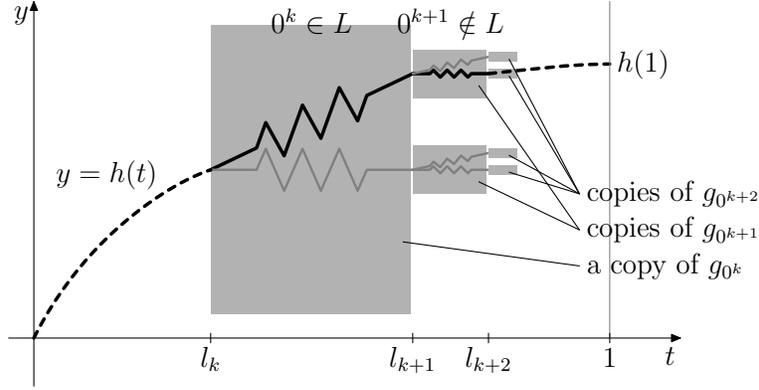}%
 \caption{%
  Compare with \ref{figure: patchwork}. 
  The blocks $g _{0 ^k}$ are now stacked vertically 
  so that the tally language~$L$ can be recovered 
  from $h (1)$. 
 }
 \label{figure: patchwork value}
\end{center}
\end{figure}

  We will construct $g$ and $h$ satisfying \kome{}
  such that the tally language $L \in \classPSPACE$ reduces to $h (1)$. 
  Apply \ref{lemma: main} to $\lambda (k) = k + 1$ 
  to obtain the polynomial~$\rho$ 
  and the families $(g _u) _u$, $(h _u) _u$. 
  As we did for \ref{theorem: main}, 
  we are going to divide $[0, 1)$ into countably many intervals $[l _k, l _{k + 1}]$
  and put there some copies of the block $g _{0 ^k}$ of \ref{lemma: main}; 
  but this time, we do not put the mirror reflection to bring $h$ back to $0$
  (\ref{figure: patchwork value}). 
  The values $h _u (1)$ pile up, 
  so that we can recover any of them by looking at $h (1)$. 

  Since $(g _u) _u$ is polynomial-time computable, 
  there is a monotone polynomial $
\gamma \tcolon \Nset \tto \Nset 
  $ satisfying $
\lvert g _{0 ^k} (t, y) \rvert \leq 2 ^{\gamma (k)}
  $ for each $k$. 
  Let $l _k = 1 - 2 ^{-k}$ and define 
\begin{gather}
 \label{equation: goal}
  g \biggl( 
   l _k + \frac{t}{2 ^{k + 1}},
    \frac{2 j + (-1) ^j y}{2 ^{2 k + \gamma (k) + \overline \rho (k)}}
  \biggr)
 = 
   \frac{g _{0 ^k} (t, y)}{2 ^{k - 1 + \gamma (k) + \overline \rho (k)}},
\\
 \label{equation: vert}
  h \biggl( 
   l _k + \frac{t}{2 ^{k + 1}} 
  \biggr)
 = 
   \frac{h _{0 ^k} (t)}{2 ^{2 k + \gamma (k) + \overline \rho (k)}}
  + 
   \sum _{\kappa = 0} ^{k - 1}
    \frac{L (0 ^\kappa)}{2 ^{2 \kappa + \gamma (\kappa) + \overline \rho (\kappa) + \rho (\kappa)}}
\end{gather}
  for each $k \in \Nset$, 
  $t \in [0, 1]$, $y \in [-1, 1]$ and $j \in \Zset$, 
  where $\overline \rho (k) = \rho (0) + \dots + \rho (k - 1)$. 
  Complete the definition by $g (1, y) = 0$ and
\begin{equation}
 \label{equation: right end}
 h (1) 
= 
 \sum _{k = 0} ^\infty
  \frac{L (0 ^k)}{2 ^{2 k + \gamma (k) + \overline \rho (k) + \rho (k)}}. 
\end{equation}
  By \ref{equation: right end}, 
  the language~$L$ reduces to $h (1)$. 

  We show that these $g$ and $h$ satisfy \kome{}. 
  Well-definedness and Lipschitz continuity of $g$
  follow from 
  \short\ref{enumi: boundary} and \short\ref{enumi: Lipschitz} of 
  \ref{lemma: main}, 
  similarly to the proof of \ref{theorem: main}. 
  Polynomial-time computability also follows from 
  that of $(g _u) _u$ again. 
  Since 
  all terms under the summation symbol in \ref{equation: vert} 
  are divisible by $
4 / 2 ^{2 k + \gamma (k) + \overline \rho (k)}
  $, 
  substituting \ref{equation: vert}
  into the second argument of \ref{equation: goal} yields 
\begin{equation}
  g \biggl( 
   l _k + \frac{t}{2 ^{k + 1}},
   h \biggl( l _k + \frac{t}{2 ^{k + 1}} \biggr)
  \biggr)
 = 
   \frac{g _{0 ^k} \bigl( t, h _{0 ^k} (t) \bigr)}{2 ^{k - 1 + \gamma (k) + \overline \rho (k)}}
 = 
   \frac{h' _{0 ^k} (t)}{2 ^{k - 1 + \gamma (k) + \overline \rho (k)}}
 =
  h' \biggl( 
   l _k + \frac{t}{2 ^{k + 1}} 
  \biggr), 
\end{equation}
  where the second and third equalities are by 
  \short\ref{enumi: equation} and \ref{equation: vert}, 
  respectively. 
  This and $h' (1) = 0 = g (h (1))$ yield \ref{equation: problem}. 
  We have proved \kome{}. 

\subsection{Proof of \ref{theorem: main exp}}
 \label{subsection: proof of theorem main exp}

\begin{figure}
\begin{center}
 \includegraphics[clip,scale=.91]{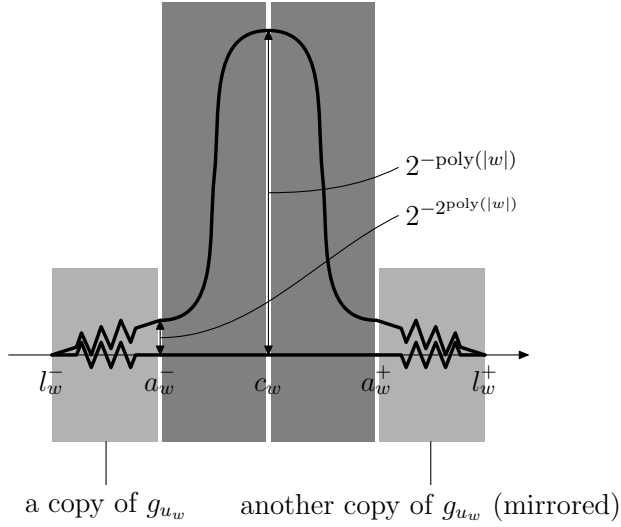}%
 \caption{%
  The construction for \ref{theorem: main exp}
  is similar to \ref{figure: patchwork}, 
  but this time we put an ``amplifier'' in the middle. 
  Then $w \in L$ if and only if 
  $h (c _w)$ is positive, 
  and in this case it is only polynomially small in $\lvert w \rvert$. 
 }
 \label{figure: patchwork exp}
\end{center}
\end{figure}

Let $L \in \classPSPACE$ be the set of triples $(M, x, 0 ^s)$ such that 
$M$ encodes a Turing machine that, 
on string input~$x$, uses at most $s$ tape cells and accepts. 
For each triple $w = (M, x, s)$, let $u _w = (M, x, 0 ^s)$. 
With a suitable encoding, we have $\lvert u _w \rvert \leq 2 ^{\lvert w \rvert}$. 
It is easy to see that $L' = \{\, w : u _w \in L \,\}$ is $\classEXPSPACE$-complete. 

The desired $g$ and $h$ such that $L'$ reduces to $h$
will be constructed as follows. 
As we did for \ref{theorem: main}, 
we divide $[0, 1)$ into infinitely many intervals $[l ^- _w, l ^+ _w]$
with midpoints $c _w$, 
and put there the functions $g _{u _w}$ of \ref{lemma: main}
to compute whether $u _w \in L$, which is equivalent to $w \in L'$. 
But this time, the outcome $h _{u _w} (1)$ is 
exponentially small in $\lvert w \rvert$ (even when it is positive), 
so we need ``amplifiers'' 
to make the value $h (c _w)$ visibly large
(\ref{figure: patchwork exp}). 
Because we use stronger and stronger amplifiers as $\lvert w \rvert \to \infty$, 
the function~$g$ will not satisfy 
the full Lipschitz condition~\ref{equation: Lipschitz}, 
but it still satisfies the weaker condition~\ref{equation: local Lipschitz}. 

Now we fill in the details. 
Apply \ref{lemma: main} to $\lambda (k) = 0$ (and the above $L$)
to obtain the polynomial~$\rho$ and families $(g _u) _u$, $(h _u) _u$. 
Since $(g _u) _u$ is polynomial-time computable, 
there is a polynomial $\gamma$ 
satisfying $
\lvert g _u (t, y) \rvert \leq 2 ^{\gamma (\lvert u \rvert) - \lvert u \rvert}
$.  We may assume that $\gamma (k) - k$ is strictly increasing in $k$ and that $
 (1.5 \ln 2) \gamma (k) 
\leq 
 2 ^{\gamma (k) - k}
$ for all $k$. 
For each binary string~$w$, 
let $\varLambda _w = 2 ^{2 \lvert w \rvert + 3}$, 
$\varGamma _w = 2 ^{\gamma (2 ^{\lvert w \rvert})}$ and 
\begin{align}
  c _w 
&
 =
   1
  -
   \frac{1}{2 ^{\lvert w \rvert}} 
  + 
    \frac{4 \overline w + 2}{\varLambda _w}, 
&
  a ^\mp _w
&
 =
  c _w \mp \frac{1}{\varLambda _w}, 
&
  l ^\mp _w
&
 =
  c _w \mp \frac{2}{\varLambda _w}, 
\end{align}
  where $\overline w \in \{0, \dots, 2 ^{\lvert w \rvert} - 1\}$ is $w$ read as an integer in binary notation. 
  This divides $[0, 1)$ into intervals $[l ^- _w, l ^+ _w]$
  of length $4 / \varLambda _w$
  indexed by $w \in \{0, 1\} ^*$. 
  Define
\begin{align}
 \label{equation: vector exp}
  g \biggl( 
   l ^\mp _w \pm \frac{t}{\varLambda _w},
   \frac{y}{\varLambda _w \varGamma _w} 
  \biggr)
&
 = 
  \pm \frac{g _{u _w} (t, \Hat y)}{\varGamma _w}, 
\\
 \label{equation: magnifier vector exp}
  g \biggl( 
   a ^\mp _w \pm \frac{t}{\varLambda _w},
   Y
  \biggr)
&
 = 
  \pm 6 t (1 - t) Y \varLambda _w \ln \varGamma _w, 
\\ % \intertext{and}
 \label{equation: flow exp}
  h \biggl( 
   l ^\mp _w \pm \frac{t}{\varLambda _w}
  \biggr)
&
 = 
  \frac{h _{u _w} (t)}{\varLambda _w \varGamma _w}, 
\\
 \label{equation: magnifier flow exp}
  h \biggl( 
   a ^\mp _w \pm \frac{t}{\varLambda _w}
  \biggr)
&
 = 
  \frac{\varGamma _w ^{t ^2 (3 - 2 t)} h _{u _w} (1)}{\varLambda _w \varGamma _w} 
\end{align}
for each $t \in [0, 1]$ and $y$, $Y \in \Rset$, 
where $\Hat y = \max \{-1, \allowbreak \min \{1, y\}\}$. 
Let $g (1, Y) = h (1) = 0$ for each $Y \in \Rset$. 
This defines $g$ and $h$ seamlessly by 
\ref{enumi: boundary}. 
Recall the idea explained in \ref{figure: patchwork exp}: 
the equations \ref{equation: vector exp} and \ref{equation: flow exp}
are analogous to \ref{equation: vector} and \ref{equation: flow}
in the proof of the main theorem, 
and \ref{equation: magnifier vector exp} and \ref{equation: magnifier flow exp}
stand for the magnifier. 

We show that $g$ and $h$ satisfy \kome{}
with \ref{equation: Lipschitz} replaced by \ref{equation: local Lipschitz}. 

  For the equation~\ref{equation: problem}, 
  it is again easy to see that $h (0) = 0$ and $h' (1) = 0 = g (1, h (1))$. 
  Numbers in $[0, 1)$ can be written 
  either as $l ^\mp _w \pm t / \varLambda _w$ or as $a ^\mp _w \pm t / \varLambda _w$, 
  and for them the equation follows respectively by 
\begin{align}
 \label{equation: derivative exp}
&
  h' \biggl( 
    l ^\mp _w \pm \frac{t}{\varLambda _w}
  \biggr)
 =
  \pm \frac{h' _{u _w} (t)}{\varGamma _w}
 =
  \pm \frac{g _{u _w} (t, h _{u _w} (t))}{\varGamma _w}
\\
\notag
& \qquad
 =
  g \biggl( 
   l ^\mp _w \pm \frac{t}{\varLambda _w},
   \frac{h _{u _w} (t)}{\varLambda _w \varGamma _w} 
  \biggr)
 =
  g \biggl( 
   l ^\mp _w \pm \frac{t}{\varLambda _w},
   h \biggl( 
    l ^\mp _w \pm \frac{t}{\varLambda _w}
   \biggr)
  \biggr), 
\\
 \label{equation: magnifier derivative exp}
&
  h' \biggl( 
    a ^\mp _w \pm \frac{t}{\varLambda _w}
  \biggr)
 =
  \pm \frac{6 t (1 - t) \ln \varGamma _w}{\varGamma _w} \varGamma _w ^{t ^2 (3 - 2 t)} h _{u _w} (1)
\\
\notag
& \qquad
 =
  g \biggl( 
   a ^\mp _w \pm \frac{t}{\varLambda _w},
   \frac{\varGamma _w ^{t ^2 (3 - 2 t)} h _{u _w} (1)}{\varLambda _w \varGamma _w} 
  \biggr)
 =
  g \biggl( 
   a ^\mp _w \pm \frac{t}{\varLambda _w},
   h \biggl( 
    a ^\mp _w \pm \frac{t}{\varLambda _w}
   \biggr)
  \biggr), 
\end{align}
  where we used 
  \ref{equation: flow exp}, 
  \short\ref{enumi: equation}, 
  \ref{equation: vector exp}, 
  \ref{equation: flow exp}, 
  \ref{equation: magnifier flow exp}, 
  \ref{equation: magnifier vector exp}, 
  \ref{equation: magnifier flow exp}
  for each equality. 

  The condition~\ref{equation: local Lipschitz} is satisfied
  with $r (k) = 2 k + 3 + s (k)$, 
  where $s$ is any polynomial such that 
  $2 ^{s (k)} \geq (1.5 \ln 2) \gamma (2 ^k)$. 
  For if $T \in [l ^- _w, a ^- _w]$ or $T \in [a ^+ _w, l ^+ _w]$ for some $w$, then 
  by \short\ref{enumi: Lipschitz} and \ref{equation: vector exp}, 
  we have 
\begin{equation}
 \frac{\lvert g (T, Y _0) - g (T, Y _1) \rvert}{\lvert Y _0 - Y _1 \rvert}
\leq
 \frac{2 ^{-\lambda (\lvert u _w \rvert)}}{\varGamma _w} \varLambda _w \varGamma _w
\leq
 \varLambda _w
=
 2 ^{2 \lvert w \rvert + 3}
\leq 
 2 ^{r (\lvert w \rvert)}. 
\end{equation}
  If $T \in [a ^- _w, a ^+ _w]$ for some $w$, then 
  by \ref{equation: magnifier vector exp} we have 
\begin{multline}
 \frac{\lvert g (T, Y _0) - g (T, Y _1) \rvert}{\lvert Y _0 - Y _1 \rvert} 
\leq
 2 ^{2 \lvert w \rvert + 3} \cdot 1.5 \ln \varGamma _w 
\\
=
 2 ^{2 \lvert w \rvert + 3} \cdot 1.5 \gamma (2 ^{\lvert w \rvert}) \ln 2
\leq
 2 ^{2 \lvert w \rvert + 3} \cdot 2 ^{s (\lvert w \rvert)} 
=
 2 ^{r (\lvert w \rvert)}. 
\end{multline}

  To see that $g$ is polynomial-time computable, 
  we use the characterization in 
  (the obvious two-dimensional generalization of) 
  \ref{lemma: polynomial modulus}. 
  Suppose we are asked for 
  an approximation of $g (T, Y)$
  to precision~$2 ^{-m}$
  for some $T \in [0, 1] \cap \Qset$ and $Y \in [-1, 1] \cap \Qset$. 
  We first find a string $w$ and $t \in [0, 1] \cap \Qset$ 
  such that $T$ can be written as $l ^\mp _w \pm t / \varLambda _w$ 
  or as $a ^\mp _w \pm t / \varLambda _w$. 
  In the latter case, 
  it is easy to compute the desired approximation 
  using \ref{equation: magnifier vector exp}. 
  In the former case, we use \ref{equation: vector exp} as follows: 
\begin{itemize}
 \item 
  If $m < 2 ^{\lvert w \rvert}$, we can safely answer $0$, 
  because the value \ref{equation: vector exp}
  is in $
[-2 ^{-2 ^{\lvert w \rvert}}, 2 ^{-2 ^{\lvert w \rvert}}]
  $ by $
 \lvert g _{u _w} (t, \Hat y) \rvert 
\leq
 2 ^{\gamma (\lvert u _w \rvert) - \lvert u _w \rvert}
\leq
 2 ^{\gamma (2 ^{\lvert w \rvert}) - 2 ^{\lvert w \rvert}}
  $.  
 \item 
  Otherwise, we
  compute $\Hat y \in \Qset$, where $y = \varLambda _w \varGamma _w Y$, 
  and then get the desired approximation of \ref{equation: vector exp}
  by computing $g _{u _w} (t, \Hat y)$ to an appropriate precision. 
  This can be done, by the polynomial-time computability of $(g _u) _u$, 
  in time polynomial in $m$ and $\lvert u _w \rvert$. 
  But this is in fact polynomial in $m$, 
  since $\lvert u _w \rvert \leq 2 ^{\lvert w \rvert} \leq m$. 
\end{itemize}
  We have thus proved \kome{}
  with \ref{equation: Lipschitz} replaced by \ref{equation: local Lipschitz}. 
  Since \ref{equation: magnifier flow exp} yields
\begin{equation}
 \label{equation: output exp}
 h (c _w) 
=
 \frac{h _{u _w} (1)}{\varLambda _w} 
=
 \frac{L (u _w)}{2 ^{2 \lvert w \rvert + 3 + \rho (\lvert w \rvert)}}, 
\end{equation}
  the language $L'$ reduces to $h$. 
  Since $L'$ is $\classEXPSPACE$-complete, 
  $h$ is $\classEXPSPACE$-hard. 

  Finally, 
  we claim that $h$ has a polynomial modulus of continuity. 
  Precisely, 
  we show that $\lvert h (T _0) - h (T _1) \rvert < 2 ^{-k}$
  whenever $k \in \Nset$ and $0 \leq T _1 - T _0 < 2 ^{-s (k) - k}$. 
\begin{itemize}
 \item 
  If $T _1 \geq 1 - 2 ^{-k}$, 
  then $T _1 \in [l ^- _w, l ^+ _w]$ for some $w$ of length $\geq k$, 
  so $\lvert h _1 (T _1) \rvert \leq 1 / \varLambda _w \leq 1 / 2 ^{2 k + 3}$
  by \ref{equation: flow exp}, \ref{equation: magnifier flow exp}
  and \short\ref{enumi: range}. 
  Applying the same argument to 
  $T _0 > T _1 - 2 ^{-s (k) - k} \geq 1 - 2 ^{-(k - 1)}$, 
  we obtain $
\lvert h _1 (T _0) \rvert \leq 1 / 2 ^{2 (k - 1) + 3}
  $.  Thus, $
 \lvert h (T _0) - h (T _1) \rvert
\leq
 \lvert h (T _0) \rvert + \lvert h (T _1) \rvert
\leq
 1 / 2 ^{2 k + 3} + 1 / 2 ^{2 (k - 1) + 3}
< 
 2 ^{-k}
  $.  
 \item 
  If $T _1 \leq 1 - 2 ^{-k}$, then 
  each point $T \in [T _0, T _1]$
  belongs to $[l ^- _w, l ^+ _w]$ for some $w$ of length $< k$. 
  If $T \in [l ^- _w, a ^- _w] \cup [a ^+ _w, l ^+ _w]$, then $
 \lvert h' (T) \rvert
\leq
 2 ^{\gamma (\lvert u _w \rvert)} / \varGamma _w
\leq
 1
  $ by the first line of \ref{equation: derivative exp}; 
  otherwise, $
 \lvert h' (T) \rvert
\leq
 6 (1 / 4) (\ln \varGamma _w) h _{u _w} (1)
=
 (1.5 \ln 2) \gamma (2 ^{\lvert w \rvert}) h _{u _w} (1)
\leq 
 2 ^{s (\lvert w \rvert)} 
\leq
 2 ^{s (k)}
  $ by the first line of \ref{equation: magnifier derivative exp}. 
  We thus have $\lvert h' (T) \rvert \leq 2 ^{s (k)}$, 
  and hence $
 \lvert h (T _0) - h (T _1) \rvert 
\leq
 2 ^{s (k)} (T _1 - T _0) 
<
 2 ^{s (k)} \cdot 2 ^{-s (k) - k}
=
 2 ^{-k}
  $. 
\end{itemize}

\section{Related results and open problems}
 \label{section: related}

\subsection{Other results on differential equations}

  \ref{table: results on initial value problems} 
  summarizes what is known about the 
  computability and complexity of the initial value problem~\ref{equation: problem}
  in our sense. 
  Computability of other aspects of the solution 
  is discussed by 
\cite{cenzer04:_index_sets_for_comput_differ_equat}, 
\cite{graca08:_bound_of_domain_of_defin} and
\cite{kawamura09:_differ_recur}. 
\cite{edalat04:_domain_theor_accoun_of_picar_theor}
  give a domain-theoretic account for the problem. 

\begin{table}[b!]
\begin{center}
 \caption{Assuming that $g$ is polynomial-time computable, 
          how complex can the solution~$h$ of~\ref{equation: problem} be?}
 \label{table: results on initial value problems}
\small
\def\arraystretch{1.7}
\newcommand{\alength}{70pt}
\newcommand{\blength}{139pt}
\newcommand{\clength}{216pt}
\begin{tabular}{p{\alength}|p{\blength}|p{\clength}}
 Assumptions & Upper bounds & Lower bounds \\ \hline
 None & ------ & \parbox[t]{\clength}{\raggedright can be (non-unique and) all non-computable: \cite{aberth71:_failur_in_comput_analy_of}, \cite{pour-el79:_comput_ordin_differ_equat_which}, \cite{ko83:_comput_compl_of_ordin_differ_equat}} \\
 $h$ is the unique solution & \parbox[t]{\blength}{computable: \cite{osgood98:_beweis_exist_einer_l_osung}, \cite{pour-el79:_comput_ordin_differ_equat_which}} & \parbox[t]{\clength}{can take arbitrarily long time: \cite{miller70:_recur_funct_theor_and_numer_analy}, \cite{ko83:_comput_compl_of_ordin_differ_equat}} \\
 condition~\ref{equation: local Lipschitz} & \parbox[t]{\blength}{exponential-space: \cite{ko92:_comput_compl_of_integ_equat}} & \parbox[t]{\clength}{can be $\classEXPSPACE$-hard (our \ref{theorem: main exp})} \\
 \parbox[t]{\alength}{the Lipschitz\\condition~\ref{equation: Lipschitz}}  & \parbox[t]{\blength}{polynomial-space: \cite{ko83:_comput_compl_of_ordin_differ_equat}} & \parbox[t]{\clength}{can be $\classPSPACE$-hard (our \ref{theorem: main})} \\
 $g$ is analytic & \parbox[t]{\blength}{polynomial-time:\\\cite{ko88:_comput_power_series_in_polyn_time},\\\cite{kawamura:_analy_initial_value_probl_in_polyn_time}} & ------
\end{tabular}
\end{center}
\end{table}

  The status between the last two rows of 
  \ref{table: results on initial value problems} remains open. 
  What happens, for example, if we assume that $g$ is infinitely often differentiable? 

  Computability (or not) of other classes of differential equations
  is studied by 
\cite{pour-el81:_wave_equat_with_comput_initial}, 
\cite{pour-el97:_wave_equat_with_comput_initial}, 
\cite{gay01:_comput_of_solut_of_kortew}, 
\cite{weihrauch02:_is_wave_propag_comput_or}, 
%      weihrauch05:_comput_solut_of_kortew, 
\cite{weihrauch06:_comput_schr_oding_propag_type_turin_machin} and
\cite{zhong07:_comput_analy_of_bound_value}. 
  Less is known about their computational complexity. 

\subsection{Constructive versions}

  A reasonable criticism about the results in 
  \ref{table: results on initial value problems} is that 
  they deal with the complexity of each single solution~$h$, 
  whereas the practical concern for numerical analysts 
  would be the complexity of the \emph{operator} that ``computes $h$ from $g$.'' 
  For computability, such constructive formulation is possible 
  through a suitable representation of the function space 
  (see Chapter~6 of 
\cite{weihrauch00:_comput_analy}). 
  For complexity, formulation of constructive results requires some more ideas, 
  and is undertaken in a recent work of 
\cite{kawamura_stoc}. 

\begin{acknowledge}
The author is grateful to Stephen A.~Cook, Ker-I Ko 
and the anonymous referees 
for comments on the manuscript 
that helped improve the presentation. 

This work was supported in part 
by the Nakajima Foundation and 
by the Natural Sciences and Engineering Research Council of Canada.

A preliminary version of this work appeared as 
\cite{kawamura09:_lipsc_contin_ordin_differ_equat}. 
\end{acknowledge}

\end{document}